\def\etcl{$\kappa$-(BE\-DT\--TTF)$_2$\-Cu\-[N\-(CN)$_{2}$]Cl}
\def\etcn{$\kappa$-(BE\-DT\--TTF)$_2$\-Cu$_2$(CN)$_{3}$}
\def\EtMe{EtMe$_3$\-Sb\-[Pd(dmit)$_2$]$_2$}

\def\cm{cm$^{-1}$}
\documentclass[aps,prl,twocolumn,showpacs,superscriptaddress,am]{revtex4}
\usepackage[dvips]{graphicx}
\usepackage{amsmath}
\usepackage{amssymb}

\begin{document} 
\title{Low-Energy Excitations in the
Quantum Spin-Liquid $\kappa$-(BE\-DT\--TTF)$_2$\-Cu$_2$(CN)$_{3}$}
\author{A. Pustogow}
\affiliation{1.~Physikalisches Institut, Universit\"{a}t
Stuttgart, Pfaffenwaldring 57, D-70550 Stuttgart Germany}
\author{E.~Zhukova}
\affiliation{1.~Physikalisches Institut, Universit\"{a}t
Stuttgart, Pfaffenwaldring 57, D-70550 Stuttgart Germany}
\affiliation{A.M. Prokhorov General Physics Institute, Russian Academy of Sciences,
119991 Moscow, Russia}
\affiliation{Moscow Institute of Physics and Technology (State University), 141700, Dolgoprudny, Moscow Region, Russia}
\author{B. Gorshunov}
\affiliation{1.~Physikalisches Institut, Universit\"{a}t Stuttgart,
Pfaffenwaldring 57, D-70550 Stuttgart Germany}
\affiliation{A.M. Prokhorov General Physics Institute, Russian Academy of Sciences,
119991 Moscow, Russia}
\affiliation{Moscow Institute of Physics and Technology (State University), 141700, Dolgoprudny, Moscow Region, Russia}
\author{M.\ Pinteri\'{c}}
\affiliation{Institut za fiziku, P.O.Box 304, HR-10001 Zagreb, Croatia}
\affiliation{Faculty of Civil Engineering, Smetanova 17, 2000 Maribor, Slovenia}
\author{S.~Tomi\'{c}}
\affiliation{Institut za fiziku, P.O.Box 304, HR-10001 Zagreb, Croatia}
\author{J. A. Schlueter}
\affiliation{Material Science Division, Argonne National Laboratory,
Argonne, Illinois 60439-4831, U.S.A.}
\author{M. Dressel}
\affiliation{1.~Physikalisches Institut, Universit\"{a}t
Stuttgart, Pfaffenwaldring 57, D-70550 Stuttgart Germany}
\date{\today}
\begin{abstract}
The electrodynamic response of the organic spin-liquid candidate $\kappa$-(BEDT-TTF)$_2$Cu$_2$(CN)$_3$ has been measured in
an extremely wide energy range ($10^{-13}$ to 2 eV)
as a function of temperature (5 to 300~K).
Below the Mott gap, excitations from the un-gapped spinon continuum cause a considerable contribution to the infrared conductivity, as suggested by the U(1) gauge theory. At THz frequencies we can identify a power-law behavior $\sigma(\omega) \propto  \omega^{\beta}$ with two distinct exponents $\beta$ that change from 0.9 to 1.3 at low temperatures.
The corresponding crossover scales with temperature: $\hbar\omega_c \approx k_B T$. The observed exponents differ by more than a factor of 2 from the theoretically predicted ones.
The findings are compared with those obtained on Herbertsmithites.
\end{abstract}


\maketitle
%
%

Quantum spin-liquids are an intriguing state of matter  \cite{Lee08,Balents10}:
although the spins interact rather strongly,
quantum fluctuations prevent long-range magnetic order
down to lowest temperatures
even in two and three dimensions.
It took decades before the theoretical concept \cite{Pomeranchuk41,Anderson73}
was realized,
first in the organic compounds \etcn\ and \EtMe, which crystallize in a triangular pattern \cite{Shimizu03,Itou08},
and later in the kagome lattice of ZnCu$_3$\-(OH)$_6$Cl$_2$
\cite{Shores05,Mendels07,Helton07}. Still the smoking gun experiment is lacking, and
also the theoretical description of real systems is unsatisfactory; by now
the nature of spin liquid state is rather unclear.

The Herbertsmithite ZnCu$_3$(OH)$_6$Cl$_2$ is a layered antiferromagnetic insulator with no magnetic order above 50~mK \cite{Mendels07,Imai08}. 
Inelastic neutron scatting experiments did not find
indications of a spin gap down to 0.1~meV, inferring that the spin excitations form a continuum \cite{Helton07,Han12}. This important issue, however, is far from being settled
neither from the experimental nor from the theoretical side \cite{Yan11,Potter13}.
Among other suggestions \cite{Kalmeyer87,Wen02,Sheng09,Qi09,Mishmash13},
it was proposed that a gapless U(1) spin-liquid state forms with a spinon Fermi surface  and  with a Dirac fermions excitation spectrum of the kagome lattice \cite{Motrunich05,Ran07}.

The situation is rather similar to the organic systems
where the molecular dimers with spin~$\frac{1}{2}$ form a highly frustrated triangular lattice  \cite{Kandpal09,Nakamura09}.
At ambient pressure no indication of N{\'e}el order is observed at temperatures as low as 20~mK, despite the considerable antiferromagnetic exchange of $J\approx 220-250$~K \cite{Shimizu03,Itou08}. The origin of the spin-liquid phase is unresolved since pure geometrical frustration should not be sufficient to stabilize the quantum spin-liquid state \cite{Huse88,Capriotti99,Kaneko14}.
In contrast to the completely insulating Herbertsmithites, where the on-site Coulomb repulsion $U$ is  large compared to the hopping integral $t$, the charge transfer salt \etcn\ is close to crossover from a Mott insulator to a metal ($U\approx t$); in fact it becomes superconducting at $T_c=3.6$~K under hydrostatic pressure of only 4~kbar \cite{Geiser91,Shimizu03}.
From heat capacity measurements gapless spin excitations have been concluded \cite{Yamashita08}
in contrast to thermal transport data \cite{Yamashita09}.
Based on a U(1) gauge theory of the Hubbard model, Lee and collaborators \cite{Lee05} suggested that the spin excitations in \etcn\ exhibit a Fermi surface and show up in the thermal conductivity.
Due to the coupling with the internal gauge field, the spinons at the Fermi surface may contribute to the optical conductivity \cite{Ioffe89,Ng07,Potter13} and the magneto-optical Faraday effect \cite{Colbert14}.
Investigations of the electrodynamic behavior may help to
establish the spin-liquid ground state and elucidate the low-energy elementary excitations.

All spin-liquid candidates are Mott insulators and the charge excitations are gapped
by the more or less large $U$.
In general, optical experiments are not sensitive to spinon excitations; however, spin-charge interactions may contribute to the low-energy optical conductivity through the emergent gauge field in the U(1) spin-liquid state \cite{Ng07,Zhou13}. This would cause a power-law behavior: $\sigma_1(\omega)\propto \omega^{\beta}$ with $\beta=2$ at low frequencies and a crossover to $\beta=3.3$  above $\hbar\omega_c \approx k_BT$. Very recently, THz investigations provided
first experimental results on
the kagome crystal  ZnCu$_3$(OH)$_6$Cl$_2$ \cite{Pilon13}: at $T=4$~K they obtained a power-law behavior of the optical conductivity with an exponent $\beta = 1 - 2$ up to  1.4~THz when it crosses over into a phonon tail.

\begin{figure}
\centering
\includegraphics[width=\columnwidth]{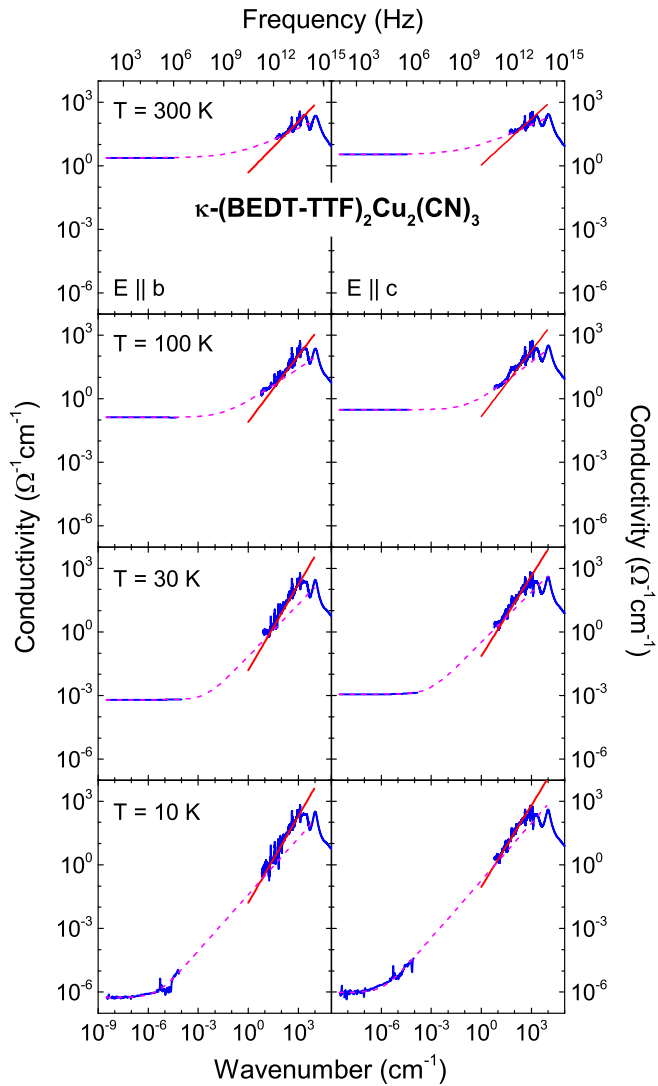}
\caption{(Color online) Overall optical conductivity of \etcn\ for the two polarizations $E\parallel b$ (left column) and $E\parallel c$ (right column) for several temperatures as indicated. The dashed magenta lines are fits
by Eq.~(\ref{eq:power})
to connect the two ranges, extending to a power-law behavior in the THz range.
The solid red lines indicate the high-frequency power law.
\label{fig:Conductivity}
}
\end{figure}

For the triangular organic compounds investigations with a similar scope have not been performed up to now.
Ng and Lee \cite{Ng07} analyzed the infrared spectra of \etcn, but
the underlying data  \cite{Kezsmarki06} did not extend to low-enough frequency to allow for founded conclusions.
Also subsequent infrared reflectivity measurements by Els{\"a}sser {\it et al.} \cite{Elsasser12} could barely reach the relevant region around $\hbar \omega \approx k_BT $ down to low temperatures.
In order to provide reliable data in the decisive energy range, we have
investigated the electrodynamic response of \etcn\ over many orders of magnitude and the complete temperature range.
Most important, interferometric optical transmission measurements at THz frequencies
allowed us to  directly calculate the complex conductivity
in the most important region.
The data were supplemented by dielectric measurements in the kHz and MHz range.
We found a power-law behavior $\sigma_1(\omega)\propto \omega^{\beta}$ up to a crossover frequency $\omega_c$ above which the slope approximately doubles.
$\beta$ depends on temperature, starting with $\beta\approx 0.4$ and 0.8 at $T=300$~K  it increases to $\beta\approx 0.9$ and 1.3 upon cooling.

Large but thin single crystals (up to $0.1 \times 4.2 \times 2.6~{\rm mm}^3$) of \etcn\ are grown by electrochemical methods \cite{Geiser91}.
The temperature dependent dc transport and the dielectric measurements between 40~Hz and 10~MHz have been described and analyzed in detail in Ref.~\onlinecite{Pinteric14}.
The optical transmission in the THz spectral range from $6 - 45$~\cm\
($0.7 - 5.6$~meV) was measured by a Mach-Zehnder interferometer based on backward-wave oscillators as powerful and tunable sources of coherent radiation \cite{Gorshunov05}. The light was polarized along the $b$ and $c$ axes of the shiny crystal surface. From the independently measured transmission coefficient and phase shift the real and imaginary parts of the conductivity, $\sigma_1(\omega)+ {\rm i}\sigma_2(\omega)$, were directly calculated without using Kramers-Kronig relations \cite{DresselGruner02}.
In order to get an overall picture of the optical properties,
complementary reflectivity measurements of the same crystal were conducted using Fourier-transform infrared spectroscopy covering the range from 20 to 12\,000~\cm,
supplemented by literature values \cite{Visentini98}.
The data were analyzed the common way \cite{Faltermeier07,Elsasser12}.

\begin{figure}[b]
\centering
\includegraphics[width=0.8\columnwidth]{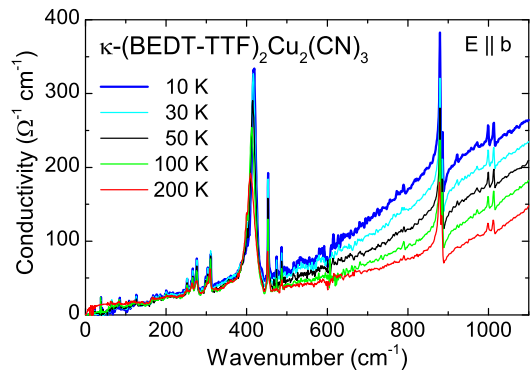}
\caption{(Color online) Temperature dependence of the far-infrared conductivity of \etcn\ for the polarization $E\parallel b$. With decreasing $T$ the in-gap absorption becomes significantly enhanced indicating non-thermal contributions to the conductivity.
\label{fig:ingap}
}
\end{figure}

In Fig.~\ref{fig:Conductivity} the conductivity $\sigma_1(\omega)$ of \etcn\ is displayed in a wide frequency range for both in-plane polarizations and different temperatures between $T=300$ and 10~K. Following the assignment developed for \etcl\  \cite{Faltermeier07,Ferber14}, the broad maximum around 2000~\cm\ can be identified as interband excitations within the BEDT-TTF dimers and excitations across the Mott gap.

\begin{figure*}
\centering
\includegraphics[width=0.9\textwidth]{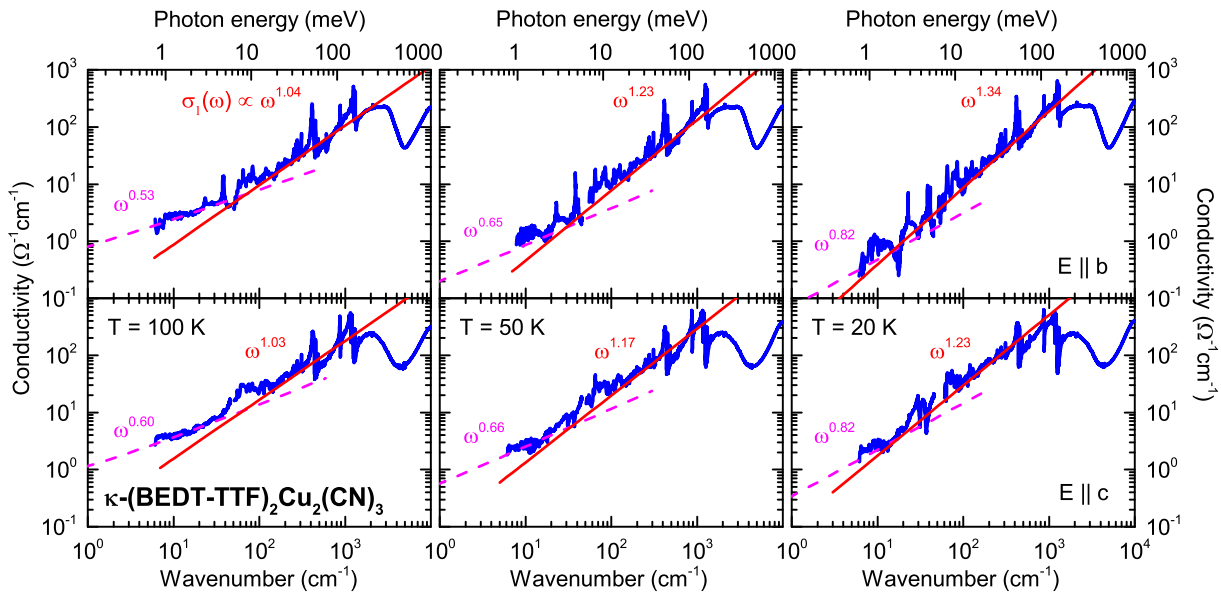}
\caption{(Color online) Optical conductivity of \etcn\ measured at $T=100$, 50, and 20~K for the two crystallographic directions, upper and lower panels as indicated. The dashed magenta lines correspond to the power-law behavior extending to low-frequencies, the solid red lines indicate the high-frequency power-laws.
\label{fig:PowerLaw}}
\end{figure*}
However, as pointed out previously \cite{Kezsmarki06,Elsasser12}, no clear-cut Mott gap is observed in the optical spectra of the spin-liquid compound \etcn; on the contrary, the absorption in the range
$100 - 1000$~\cm\ increases strongly for $T<100$~K. The extremely unusual temperature behavior is best seen in the linear plot of the optical conductivity displayed in Fig.~\ref{fig:ingap}: the far-infrared spectral weight increases significantly as the temperature drops. Note, that we do not observe a gradual filling or closing of the gap, but an increase in  the exponent $\beta$ of the power-law $\sigma(\omega)\propto \omega^{\beta}$ that is discussed in more detail in the following.
Our robust observation is an unambiguous signature that these  low-energy excitations are of quantum and not of thermal origin, strongly supporting the idea that the spin-degrees of freedom
contribute to the optical conductivity.

\begin{figure}
\centering
\includegraphics[width=0.8\columnwidth]{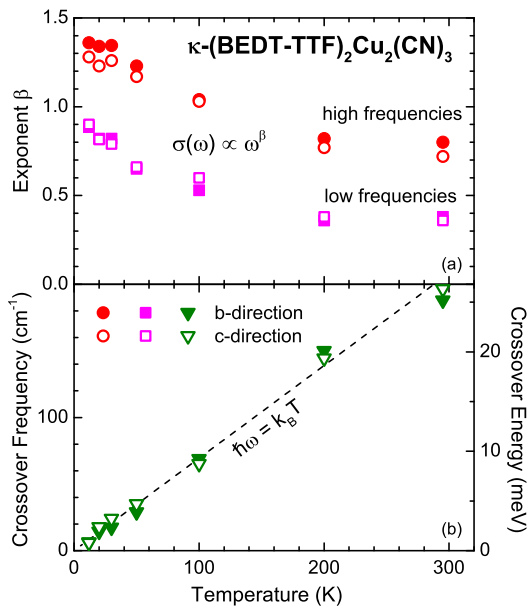}
\caption{(Color online) (a)~Temperature dependence of the power-law exponents $\sigma_1(\omega)\propto \omega^\beta$ of \etcn\
for $E\parallel b$ (solid symbols) and $E\parallel c$ (open symbols).
(b)~The crossover frequency $\omega_c$ between both regimes shifts with temperature
corresponding to $\hbar\omega_c\approx k_BT$.
\label{fig:Exponent}
}
\end{figure}

In the radio-frequency range, hopping conduction is identified as the dominant transport mechanism, accompanied by a broad dielectric relaxation at lower frequencies that bears typical fingerprints of relaxor ferroelectricity \cite{Pinteric14}. For low temperatures $T<50$~K,
we find an appreciable increase in $\sigma_1(\omega)$ of the high-frequency dielectric data, which nicely matches the slope observed in the GHz and THz range.
The dashed magenta lines in Fig.~\ref{fig:Conductivity} simply interpolate the gap in our data by
\begin{equation}
\sigma_1(\omega)=\sigma_0 + A \omega^\beta \quad ,
\label{eq:power}
\end{equation}
with a temperature-dependent constant $\sigma_0$ and prefactor $A$. The exponent $\beta$
of the power law is approximately $0.4$ and increases to almost 1 when the temperature is reduced below $T=100$~K, as summarized in Fig.~\ref{fig:Exponent}(a). Note, the rise
in $\sigma_1(\omega)$ and the corresponding power law is already observed above  300~kHz in the low-temperature dielectric data (Fig.~\ref{fig:Conductivity} lower panels). Covering such an extremely broad spectral range leads to a high confidence in the power-law exponents.

The most interesting observation, however, is the change of this power-law behavior found in the THz frequency range. In Fig.~\ref{fig:PowerLaw} the data are replotted on a magnified scale for some selected temperatures. At $T=100$~K, for example, the slope unambiguously changes around $\omega_c\approx 100$~\cm, where the exponent increases from $\beta=0.53$ to 1.04 for $E\parallel b$ (upper panel) and from 0.60 to 1.03 for $E\parallel c$ (lower panel). This crossover can be identified for all temperatures and both polarizations \cite{remark1}.
It should be noted that the high-frequency slope is not determined by a phonon tail that obscures the data in the Herbertsmithites \cite{Pilon13}.

The comprehensive results of the analysis are plotted in Fig.~\ref{fig:Exponent}(a). Below $T=100$~K the power-law exponent is enhanced reaching $\beta\approx 0.9$ and 1.36, respectively, for the lowest temperature. The difference in $\beta$ between the low and high-frequency ranges basically remains the same for all temperatures. Within the uncertainty no significant anisotropy is observed \cite{remark2}. The crossover frequency $\omega_c$ between the two regimes shifts with temperature,
as already seen from Fig.~\ref{fig:Conductivity}. Plotting $\omega_c$ for various temperatures in Fig.~\ref{fig:Exponent}(b), we find that $\hbar\omega_c\approx k_BT$ is quantitatively observed in a large temperature range.
At $T=300$~K we exceed the exchange coupling $J\approx 250$~K and significant deviations from the power-law behavior are expected \cite{Ng07}

At first glance our findings agree well with the suggestion of
Ng and collaborators \cite{Ng07,Zhou13} that un-gapped spinons at the Fermi surface
may contribute to the optical behavior of a spin liquid.
They predict a strongly enhanced conductivity
within the Mott gap compared to the two spin wave absorption in a N{\'e}el-ordered insulator.
The significant increase in conductivity at low temperatures is a robust observation \cite{Kezsmarki06,Elsasser12} that strongly supports the conclusion of gapless spin excitations.

In addition, the calculations yield a power-law absorption at low frequencies, {\it i.e}.\
for energies smaller than the exchange coupling $J$.
For very low energies ($\hbar\omega < k_BT$) the
optical conductivity $\sigma_1(\omega)\propto\omega^2$, and for
$\hbar\omega
> k_BT$ the power law should increase to $\sigma_1(\omega) \propto
\omega^{3.33}$ if scattering is negligible \cite{Ng07}.
Undoubtedly, we can for the first time identify the crossover frequency $\omega_c$ in exactly the right range; it furthermore obeys the predicted temperature dependence.
However, our extracted power-law exponents are consistently lower
by more than a factor of 2.

Here we might speculate whether this is
due to the inherent inhomogeneities in the system: when localization effects dominate,
the electronic system behaves like a disordered metal \cite{Ng07} and exhibits the corresponding frequency dependence \cite{MottDavis79,Lee85}. Furthermore, recent data of the anisotropic charge response at low frequencies suggest that \etcn\ is not a simple Mott insulator close to the metal-insulator transition \cite{Pinteric14}; rather it might be viewed as a disordered strongly correlated system close to the Mott-Anderson phase transition \cite{Byczuk05,Aguiar09}.
Since the Herbertsmithite ZnCu$_3$(OH)$_6$Cl$_2$ also exhibits an exponent in the low-frequency power-law of only $\beta\approx 1.4$ \cite{Pilon13} and the importance of
defects and impurities has been pointed out previously \cite{Mendels07,Helton07}, this strongly indicates a fundamental issue.
No doubt, further theoretical work is needed to describe the observations quantitatively.

In conclusion, we found a strong in-gap contribution to the optical conductivity in the spin-liquid compound \etcn\ that exhibits a power-law behavior $\sigma_1(\omega)\propto \omega^\beta$ down to extremely low energies. Two ranges can be clearly identified with a crossover at $\hbar\omega_c = k_BT$, where the exponent changes from $\beta=0.9$
to 1.3 in the low-temperature limit, for instance.
Our findings support the suggestion of a large spinon Fermi surface: the gapless spin excitations contribute to the optical absorption in the U(1) quantum spin liquid.

We thank P.A. Lee and R. Valenti for valuable discussions. The project was
supported by the Deutsche Forschungsgemeinschaft (DFG) and the Russian Ministry of Education and science (Program 5 top 100). M.P and S.T. acknowledge support
in part by the Croatian Science Foundation project no. IP-2013-11-1011.


\begin{thebibliography}{99}
\bibitem{Lee08}
P. A. Lee, Science {\bf 321}, 1306 (2008).
\bibitem{Balents10}
L. Balents, Nature {\bf 464}, 199 (2010).

\bibitem{Pomeranchuk41}
I. Y. Pomeranchuk, Zh. Eksp. Teor. Fiz. {\bf 11}, 226  (1941).
\bibitem{Anderson73}P. W. Anderson, Mater. Res. Bull. {\bf 8}, 153 (1973).

\bibitem{Shimizu03}Y. Shimizu, K. Miyagawa, K. Kanoda, M. Maesato, and G. Saito, Phys. Rev. Lett. {\bf 91}, 107001 (2003);
Y. Kurosaki, Y. Shimizu, K. Miyagawa, K. Kanoda, and G. Saito, Phys. Rev. Lett. {\bf 95}, 177001 (2005).
\bibitem{Itou08}
T. Itou, A. Oyamada, S. Maegawa, M. Tamura, and R. Kato, Phys. Rev. B {\bf 77}, 104413 (2008); 
T. Itou, A. Oyamada, S. Maegawa, and R. Kato, Nature Phys. {\bf 6}, 673 (2010).

\bibitem{Shores05} M. P. Shores, E. A. Nytko, B. M. Bartlett, and D. G. Nocera, J. Am. Chem. Soc. {\bf 127}, 13 462 (2005).
\bibitem{Helton07}
J. S. Helton, K. Matan, M. P. Shores, E. A. Nytko, B. M. Bartlett, Y. Yoshida, Y. Takano, A. Suslov, Y. Qiu, J.-H. Chung, D. G. Nocera, and Y. S. Lee, Phys. Rev. Lett. {\bf 98}, 107204 (2007).
\bibitem{Mendels07} P. Mendels, F. Bert, M. A. de Vries, A. Olariu, A. Harrison, F. Duc, J. C. Trombe, J. S. Lord, A. Amato, and C. Baines, Phys. Rev. Lett. {\bf 98}, 077204 (2007).
\bibitem{Imai08}T. Imai, E. A. Nytko, B. M. Bartlett, M. P. Shores, and D. G. Nocera, Phys. Rev. Lett. {\bf 100}, 077203 (2008).
\bibitem{Han12} T.-H. Han, J. S. Helton, S. Chu, D. G. Nocera, J. A. Rodriguez-Rivera, C. Broholm, and Y. S. Lee, Nature (London) {\bf 492}, 406 (2012);
T.-H. Han, R. Chisnell, C. J. Bonnoit, D. E. Freedman, V. S. Zapf, N. Harrison, D. G. Nocera, Y. Takano, and Y. S. Lee, arXiv:1402.2693.

\bibitem{Yan11} S. Yan, D. Huse, and S. White, Science {\bf 332}, 1173 (2011).
\bibitem{Potter13}A. C. Potter, T. Senthil, and P. A. Lee, Phys. Rev. B {\bf 87}, 245106 (2013).

\bibitem{Wen02}
X.-G. Wen, Phys. Rev. B {\bf 65}, 165113 (2002).
\bibitem{Sheng09}
D. N. Sheng, O. I. Motrunich, and M. P. A. Fisher,
Phys. Rev. B {\bf 79}, 205112 (2009).


\bibitem{Kalmeyer87}
V. Kalmeyer and R. B. Laughlin,
Phys. Rev. Lett. {\bf 59}, 2095 (1987).
G. Misguich, C. Lhuillier, B. Bernu, and C. Waldtmann,
Phys. Rev. B {\bf 60}, 1064 (1999).
\bibitem{Qi09}
Y. Qi, C. Xu, and S. Sachdev,
Phys. Rev. Lett. {\bf 102}, 176401 (2009); {\it ibid.} {\bf 102}, 189903 (2009).
\bibitem{Mishmash13}
R. V. Mishmash, J. R. Garrison, S. Bieri and C. Xu, Phys. Rev. Lett. {\bf 111}, 157203 (2013).


\bibitem{Motrunich05}
O. I. Motrunich, Phys. Rev. B {\bf 72}, 045105 (2005).
\bibitem{Ran07} Y. Ran, M. Hermele, P. A. Lee, and X.-G. Wen, Phys. Rev. Lett. {\bf 98}, 117205 (2007).

\bibitem{Kandpal09}
H. C. Kandpal, I. Opahle, Y.-Z. Zhang, H. O. Jeschke, and R. Valent{\'i},
Phys. Rev. Lett. {\bf 103}, 067004 (2009).
\bibitem{Nakamura09}
K. Nakamura, Y. Yoshimoto, T. Kosugi, R. Arita, and M. Imada, J. Phys. Soc. Jpn. {\bf 78}, 083710 (2009).


\bibitem{Huse88}
D. A. Huse and V. Elser, Phys. Rev. Lett. {\bf 60}, 2531 (1988).
\bibitem{Capriotti99}
L. Capriotti, A. E. Trumper, and S. Sorella, Phys. Rev. Lett. {\bf 82}, 3899 (1999).
\bibitem{Kaneko14}
R. Kaneko, S. Morita, and M. Imada,
J. Phys. Soc. Jpn. {\bf 83}, 093707 (2014).



\bibitem{Geiser91} 
U. Geiser, H. H. Wang, K. D. Carlson, J. M. Williams, H. A.
Charlier, J. E. Heindl, G. A. Yaconi, B. J. Love, M.W. Lathrop, J.
E. Schirber, D. L. Overmyer, J. Q. Ren, and M.-H. Whangbo, Inorg.
Chem. {\bf 30}, 2586 (1991).


\bibitem{Yamashita08}
S. Yamashita, Y. Nakazawa, M. Oguni, Y. Oshima, H. Jojiri,
K. Miyagawa, and K. Kanoda, Nature Phys. {\bf 4}, 459 (2008).
\bibitem{Yamashita09}
M. Yamashita, N. Nakata, Y. Kasahara, T. Sasaki, N. Yoneyama,
N. Kobayashi, S. Fujimoto, T. Shibauchi, and Y. Matsuda, Nature
Phys. {\bf 5}, 44 (2009).



\bibitem{Lee05}
S.-S. Lee and P.A. Lee, Phys. Rev. B {\bf 72}, 235104 (2005);
S.-S. Lee, P. A. Lee, and T. Senthil,
Phys. Rev. Lett. {\bf 98}, 067006 (2007).

\bibitem{Ioffe89}L. B. Ioffe and A. I. Larkin, Phys. Rev. B {\bf 39}, 8988 (1989).
\bibitem{Ng07}
T.-K. Ng and P. A. Lee, Phys. Rev. Lett. {\bf 99}, 156402 (2007).
\bibitem{Colbert14}
J. R. Colbert, H. D. Drew, and P. A. Lee,
Phys. Rev. B {\bf 90}, 121105 (2014).
\bibitem{Zhou13}
Y. Zhou and  T.-K. Ng,
Phys. Rev. B {\bf 88}, 165130 (2013);
Y.-F. Ma and T.-K. Ng,
arXiv: 1410.6330.


\bibitem{Pilon13} D. V. Pilon, C. H. Lui, T. -H. Han, D. Shrekenhamer, A. J. Frenzel, W. J. Padilla, Y. S. Lee, and N. Gedik, Phys. Rev. Lett. {\bf 111}, 127401 (2013).

\bibitem{Kezsmarki06}
I. Kezsm{\'a}rki, Y. Shimizu, G. Mihaly, Y. Tokura, K. Kanoda, and G. Saito, Phys. Rev. B {\bf 74}, 201101  (2006).
\bibitem{Elsasser12}S. Els{\"a}sser, D. Wu, M. Dressel, and J. A. Schlueter, Phys. Rev. B {\bf 86}, 155150 (2012).


\bibitem{Pinteric14}
M. Pinteri\'{c}, M. {\v C}ulo, O. Milat, M. Basleti\'{c}, B. Korin-Hamzi\'{c}, E. Tafra, A. Hamzi\'{c}, T. Ivek, T. Peterseim, K. Miyagawa, K. Kanoda, J. A. Schlueter, M. Dressel, and S. Tomi\'{c}, Phys. Rev. B {\bf 90}, 195139 (2014).

\bibitem{Gorshunov05}
 B. P. Gorshunov, A. Volkov, I. E. Spektor, A. S. Prokhorov, A. A. Mukhin, M. Dressel, S. Uchida, and A. Loidl, Int. J. Infrared Millimeter Waves {\bf 26}, 1217 (2005).
\bibitem{DresselGruner02}M.~Dressel and G.~Gr\"{u}ner, {\it Electrodynamics of Solids}  (Cambridge University Press, Cambridge, 2002).
\bibitem{Visentini98}
G. Visentini, M. Masino, C. Bellitto, and A. Girlando,
Phys. Rev. B {\bf 58}, 9460 (1998).

\bibitem{Faltermeier07}
D. Faltermeier, J. Barz, M. Dumm, M. Dressel, N. Drichko, B.
Petrov, V. Semkin, R. Vlasova, C. M\'eziere, and P. Batail, Phys.
Rev. B {\bf 76}, 165113 (2007);
J. Merino, M. Dumm, N. Drichko, M. Dressel, and R.H. McKenzie,
Phys. Rev. Lett. {\bf 100}, 086404 (2008);
M. Dumm, D. Faltermeier, N. Drichko, M. Dressel, C. M\'eziere, and
P. Batail, Phys. Rev. B {\bf 79}, 195106 (2009);
M. Dressel D. Faltermeier, M. Dumm, N. Drichko,
B. Petrov, V. Semkin, R. Vlasova,  C. M\'ezi{\`e}re, and P.
Batail, Physica B {\bf 404}, 541 (2009).
\bibitem{Ferber14}
J. Ferber, K. Foyevtsova, H. O. Jeschke, and R. Valenti,
Phys. Rev. B {\bf 89}, 205106 (2014).
\bibitem{remark1}
The extracted values are in accord with the preliminary analysis of
the independent experiments on a different \etcn\ single crystal
presented in Fig.~5 of Ref.~\onlinecite{Elsasser12}.
\bibitem{remark2}
We note that the low-frequency optical response becomes slightly  more anisotropic
as the temperature is reduced below $T=100$~K. This is in accord with previous
optical experiments of Els\"asser {\it et al.} \cite{Elsasser12} and
K. Itoh, H. Itoh, M. Naka, S. Saito, I. Hosako, N. Yoneyama, S. Ishihara, T. Sasaki, and S. Iwai,
Phys. Rev. Lett. {\bf 110}, 106401 (2013).
\bibitem{MottDavis79}
N.F. Mott and E.A. Davis, {\em Electronic Properties of Non-Crystalline Materials},
2nd edition (Clarendon Press, Oxford, 1979).
\bibitem{Lee85}
P.A. Lee and T.V. Ramakrishnan, Rev. Mod. Phys. {\bf 57}, 287 (1985).

\bibitem{Byczuk05}
K. Byczuk, W. Hofstetter, and D. Vollhardt,
Phys. Rev. Lett. {\bf 94}, 056404 (2005).
\bibitem{Aguiar09}
M. C. O. Aguiar, V. Dobrosavljevi{\'c}, E. Abrahams, and G. Kotliar,
Phys. Rev. Lett. {\bf 102}, 156402 (2009).






\end{thebibliography}
\end{document}